\documentclass[a4paper,11pt]{article}
\pdfoutput=1 

\usepackage{jcappub} 

\usepackage[T1]{fontenc} 

\title{\boldmath Searches for clustering in the time integrated skymap of the ANTARES neutrino telescope}

\emailAdd{fabian.schussler@cea.fr}

\author[a]{S.~Adri\'an-Mart\'inez}
\author[b]{A.~Albert}
\author[c]{M.~Andr\'e}
\author[e]{G.~Anton}
\author[a]{M.~Ardid}
\author[f]{J.-J.~Aubert}
\author[g]{B.~Baret}
\author[h]{J.~Barrios-Mart\'{\i}}
\author[i]{S.~Basa}
\author[f]{V.~Bertin}
% ...

\affiliation[a]{\scriptsize{Institut d'Investigaci\'o per a la Gesti\'o Integrada de les Zones Costaneres (IGIC) - Universitat Polit\`ecnica de Val\`encia. C/  Paranimf 1 , 46730 Gandia, Spain.}}
\affiliation[b]{\scriptsize{GRPHE - Institut universitaire de technologie de Colmar, 34 rue du Grillenbreit BP 50568 - 68008 Colmar, France}}
\affiliation[c]{\scriptsize{Technical University of Catalonia, Laboratory of Applied Bioacoustics, Rambla Exposici\'o,08800 Vilanova i la Geltr\'u,Barcelona, Spain}}
% ...
\affiliation[e]{\scriptsize{Friedrich-Alexander-Universit\"at Erlangen-N\"urnberg, Erlangen Centre for Astroparticle Physics, Erwin-Rommel-Str. 1, 91058 Erlangen, Germany}}
\affiliation[f]{\scriptsize{CPPM, Aix-Marseille Universit\'e, CNRS/IN2P3, Marseille, France}}
\affiliation[g]{\scriptsize{APC, Universit\'e Paris Diderot, CNRS/IN2P3, CEA/IRFU, Observatoire de Paris, Sorbonne Paris Cit\'e, 75205 Paris, France}}
\affiliation[h]{\scriptsize{IFIC - Instituto de F\'isica Corpuscular, Edificios Investigaci\'on de Paterna, CSIC - Universitat de Val\`encia, Apdo. de Correos 22085, 46071 Valencia, Spain}}
\affiliation[i]{\scriptsize{LAM - Laboratoire d'Astrophysique de Marseille, P\^ole de l'\'Etoile Site de Ch\^ateau-Gombert, rue Fr\'ed\'eric Joliot-Curie 38,  13388 Marseille Cedex 13, France}}
% ...

\abstract{This paper reports a search for spatial clustering of the arrival directions of high energy muon neutrinos detected by the ANTARES neutrino telescope. An improved two-point correlation method is used to study the autocorrelation of 3058 neutrino candidate events as well as cross-correlations with other classes of astrophysical objects: sources of high energy gamma rays, massive black holes and nearby galaxies. No significant deviations from the isotropic distribution of arrival directions expected from atmospheric backgrounds are observed.}

\begin{document}
\maketitle
\flushbottom

\section{Introduction}\label{sec:intro}
The key question to resolve the long standing mystery of the origin of cosmic rays is to locate their sources and understand the mechanisms that accelerate these particles up to energies orders of magnitude above the energies reached by man-made accelerators. Over the last years it has become obvious that multiple messengers will be needed to achieve this task. Fundamental particle physics processes like the production and subsequent decay of pions in interactions of high energy particles predict that the acceleration sites of cosmic rays should also be sources of high energy gamma rays and neutrinos. The detection of astrophysical neutrinos and the identification of their sources is one of the main goals of neutrino telescopes operated at the South Pole (IceCube~\cite{IceCube}), in Lake Baikal~\cite{Baikal} and in the Mediterranean Sea (ANTARES~\cite{Antares_DetectorPaper}). 

Despite significant effort, no clear signature for point-like sources of astrophysical neutrinos has been found so far~\cite{Amanda_PointSources2009, IceCube_PointSources2013, IceCube_ICRC2013_PS, ANTARES_PointSources2010, ICRC2013_ANTARESExcess, ICRC2013_PointSources}. Currently, both the spatial distribution as well as the morphologies of sources potentially emitting neutrinos in the TeV energy range are unknown. Similar to the distribution of observed sources emitting high energy gamma rays, they are supposed to be distributed very inhomogeneously throughout our cosmic neighbourhood. A significant fraction of them should be located in the Galactic disk and be spatially extended (e.g. shell-type supernova remnants). It is therefore interesting to study the intrinsic clustering of the arrival directions of neutrino candidates. 

In this analysis an improved autocorrelation method is used to this end. As no prior information about the potential sources is required biases are naturally reduced. Since it covers a large angular range, i.e. neutrino emission regions of very different sizes, this study complements the searches for point-like sources and could provide hints for underlying, yet unresolved, source morphologies and source distributions. Exploiting the expected multi-messenger signatures of potential sources the introduced method is extended to searches for correlations between the arrival directions of neutrino candidates and other classes of astrophysical objects: sources of high energy gamma rays, massive black holes and nearby galaxies.
 
\subsection{The ANTARES neutrino telescope}
\begin{figure*}[!t]
\centerline {
  \includegraphics[width=0.8\textwidth]{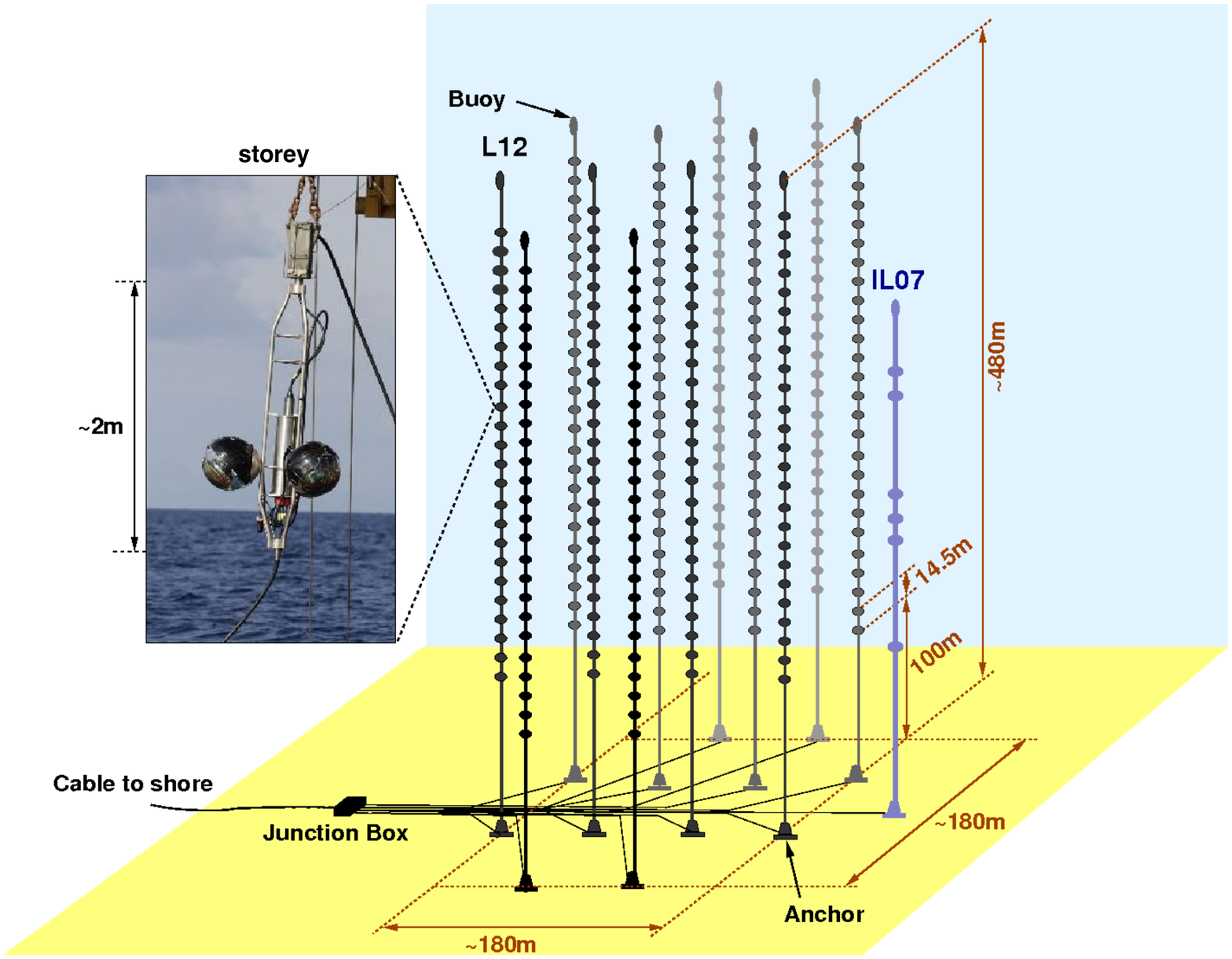}
 }     
  \caption{Schematic view of the ANTARES telescope. The inset shows a photograph of an optical storey.}
\label{fig:layout}	
 \end{figure*}
 \noindent%
 
The ANTARES telescope~\cite{Antares_DetectorPaper} became fully operational in 2008. The detector comprises twelve detection lines anchored at a depth of $2475~\mathrm{m}$ and $40~\mathrm{km}$ off the French coast near Toulon. The detector lines are about $450~\mathrm{m}$ long and host a total of 885 optical modules (OMs), each comprising a 17'' glass sphere which houses a 10'' photomultiplier tube. The OMs look downward at $45^\circ$ in order to optimise the detection of upgoing, i.e. neutrino induced, tracks. The geometry and size of the detector make it sensitive to extraterrestrial neutrinos in the TeV-PeV energy range. A schematic layout of the telescope is shown in Figure~\ref{fig:layout}.

The neutrino detection is based on the induced emission of Cherenkov light by high energy muons originating from charged current neutrino interactions inside or near the instrumented volume. All detected signals ({\it hits}) are transmitted via an optical cable to a shore station, where a computer farm filter the data for coincident signals in several adjacent OMs. The muon direction is then determined by maximising a likelihood which compares the time of the hits with the expectation from the Cherenkov signal of a muon track. Details on the event reconstruction are given in Ref.~\cite{ANTARES_PointSources2010, ANTARES_NuSpectrum}.

\begin{figure*}[!t]
\centerline {
     \includegraphics[width=0.75\textwidth]{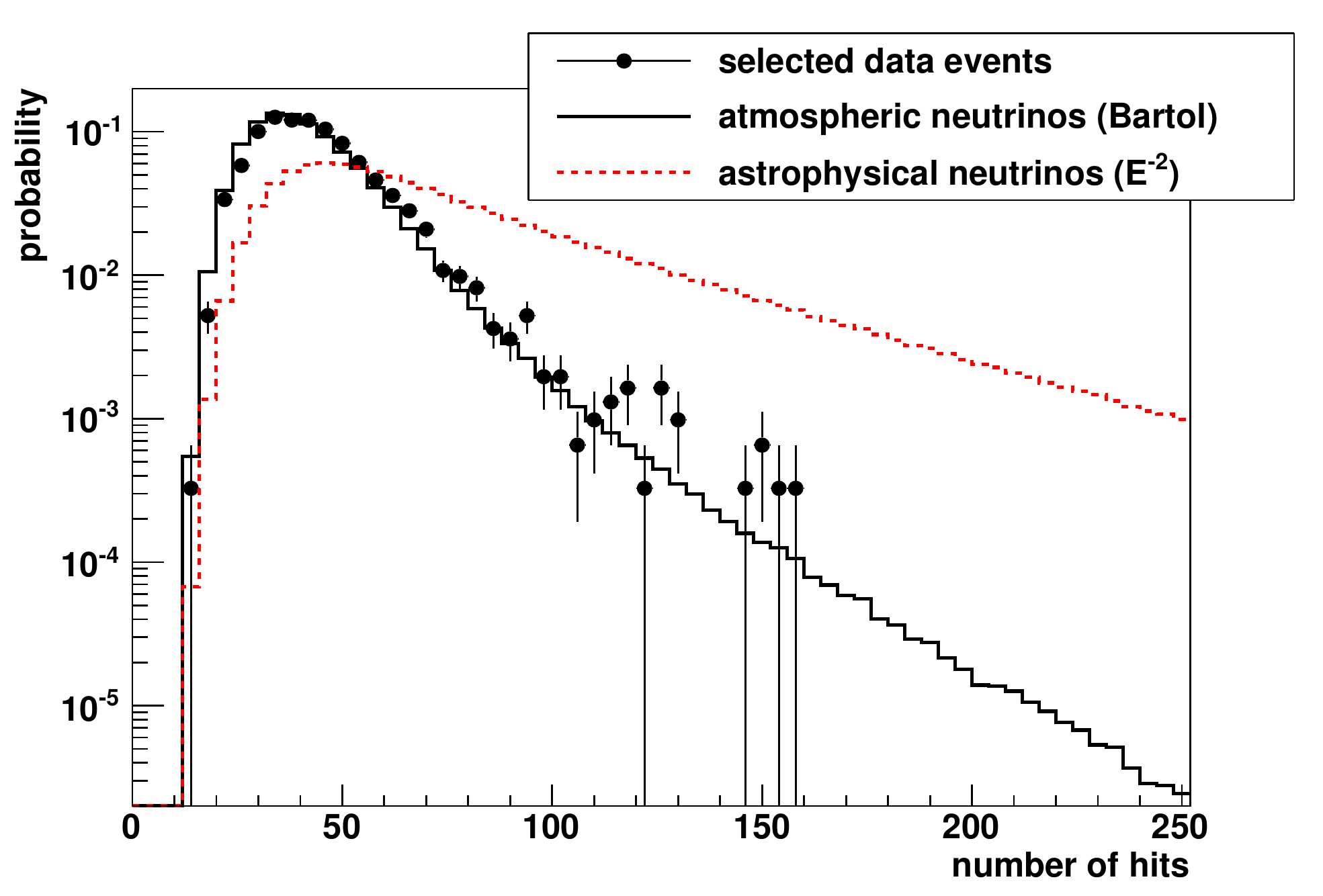}
}     
  \caption{Normalised distribution of the number of hits used in the event reconstruction for data (black markers) and Monte Carlo simulations (atmospheric neutrinos following the parametrisation from Ref.~\cite{Bartol}: black histogram; astrophysical neutrinos: red, dotted histogram).}
\label{fig:nHit}	
 \end{figure*}

Two main backgrounds for the search for astrophysical neutrinos can be identified: downgoing atmospheric muons which have been mis-reconstructed as upgoing and atmospheric neutrinos originating in cosmic ray induced air showers at the opposite side of the Earth. Depending on the requirements of the analysis both backgrounds can, at least partially, be discriminated using various parameters such as the quality of the event reconstruction or an estimator of the energy of the muon, e.g. the number of hits used in the track reconstruction. The latter, being strongly correlated with the energy of the original neutrino, helps to discriminate events of atmospheric origin from neutrinos produced in astrophysical sources. Atmospheric neutrinos have a much softer energy spectrum ($\propto E^{-3.7}$) compared to the generic $E^{-2}$ spectrum expected from Fermi acceleration in astrophysical sources. As illustrated in Figure~\ref{fig:nHit} this difference affects the distribution of energy dependent parameters (or {\it energy estimators}) and can therefore be used to enhance the background discrimination. In addition, analysing the reconstructed arrival directions of the events allows to search for an excess over the isotropic atmospheric backgrounds. Both features will be exploited in the analysis described in this paper.

 \noindent

\subsection{Dataset}
The data analysed here has been recorded by the ANTARES neutrino telescope between 2007 and 2010. During the beginning of this period (2007-2008) the detector was in its commissioning phase, increasing from five active detection lines to the full detector with twelve lines by mid 2008. After imposing basic data quality requirements, the event selection criteria have been optimised by means of Monte Carlo simulations to yield the best average upper limit on the neutrino flux in a search for point-like sources~\cite{ANTARES_PointSources2010}. These criteria are mainly a cut on the reconstructed zenith angle, $\theta > 90^\circ$, a requirement on the reconstruction quality parameter, $\Lambda$, as well as a cut on the estimated angular uncertainty of the track reconstruction, $\beta < 1^\circ$. A total of 3058 neutrino candidates are found in 813 days of effective lifetime. Monte Carlo simulations show that the contribution from misreconstructed atmospheric muons is about $15~\%$ in this dataset. A skymap in galactic coordinates of these events  is shown in the upper left plot of Figure~\ref{fig:SkyMaps}. 

 
\section{Autocorrelation analysis}\label{sec:autocorr}
\subsection{Method}\label{sec:method} 
 \begin{figure*}[!t]
\centerline {
   \includegraphics[width=0.75\textwidth]{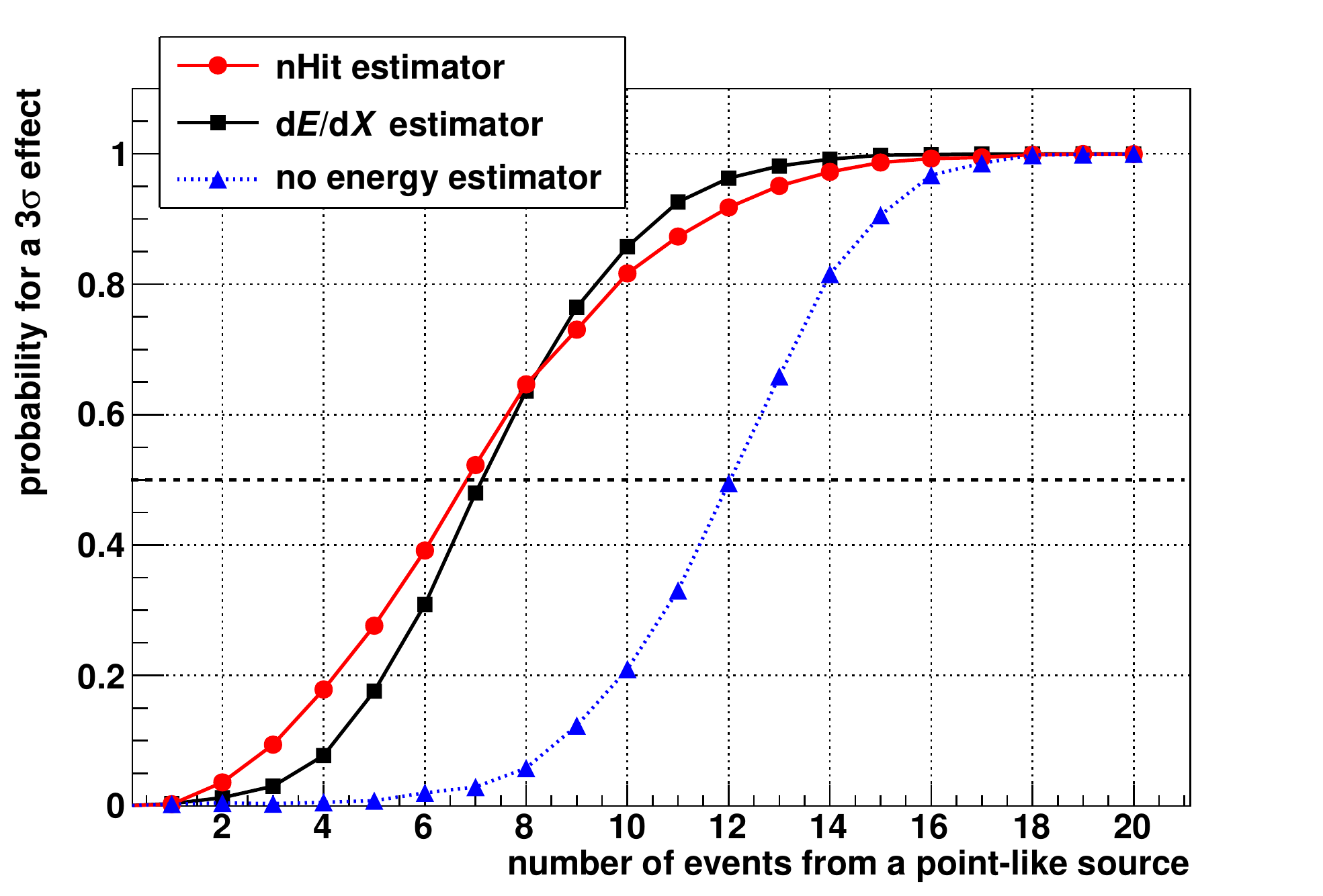}
}
\caption{Probability to detect with a $3~\sigma$ sigma significance a single point-like source following an $E^{-2}$ energy spectrum as a function of its neutrino luminosity, i.e. the number of detected neutrinos. The blue triangles denote the standard autocorrelation method without an energy estimator. The black squares show the performance including the $\mathrm{d}E/\mathrm{d}X$ energy estimator~\cite{ICRC2013_dEdX} and the red circles denote the finally used method using the $n_\mathrm{Hit}$ estimator.}
   \label{fig:Performance_Energy}
 \end{figure*}
\noindent%
The most commonly used method to detect intrinsic clusters within a set of $N$ events is the standard two-point autocorrelation distribution. It is defined as the differential distribution of the number of observed event pairs, $N_\mathrm{p}$, in the dataset as a function of their mutual angular distance, $\Delta \Omega$. This technique has already been applied to the first two years of ANTARES data. No significant clustering has been detected~\cite{ICRC2011_AutoCorrelation}. Here, an improvement of this method by using an estimator of the neutrino energy is presented. To suppress statistical fluctuations that would reduce the sensitivity of the method, the cumulative autocorrelation distribution is used. It is  defined as \begin{equation}
\mathcal{N}_{\hat{E}} (\Delta \Omega) = \sum\limits_{i=1}^{N} \sum\limits_{j=i+1}^{N} w_{ij} \cdot \left[ 1 - H(\Delta \Omega_{ij} - \Delta \Omega)\right], \label{equ:autocor}
\end{equation}
where $H$ is the Heaviside step function. The weights $w_{ij}=w_i \cdot w_j$ are calculated using the individual event weights $w_i= \int_0^{\hat{E}_i} f(\hat{E})\; \mathrm{d}\hat{E}$, where $f(\hat{E})$ is the cumulative distribution of the energy estimator $\hat{E}$ for the background. Astrophysical neutrinos are more likely to produce events with a higher value of the energy estimator $\hat{E_i}$ than atmospheric neutrinos. This is represented by a higher event weight $w_i$. The used distribution is built from large statistics Monte Carlo simulations reproducing the actual data taking conditions, including, for example, the time dependent background fluctuations induced by bioluminescence. These simulations have been validated by extensive comparisons with data. An example is shown in Figure~\ref{fig:nHit}, where the number of hits used in the event reconstruction is depicted (see~\cite{ANTARES_PointSources2010, ANTARES_NuSpectrum} for further details). The simulated events used to build the $f(\hat{E})$ distribution follow an energy spectrum as expected for atmospheric neutrinos~\cite{Bartol} (black histogram in Figure~\ref{fig:nHit}). Modifying the standard autocorrelation by these weights leads to a significant increase of the sensitivity to detect clustering of (astrophysical) events following a harder energy spectrum. This improvement is illustrated in Figure~\ref{fig:Performance_Energy}. Various possibilities exist for the estimation of the energy and the definition of the weights. As crosscheck of the stability and performance of the method, the full analysis has been performed using two different energy estimators: the number of hits used during the final step of the event reconstruction, $n_\mathrm{Hit}$, as in the search for point-like sources~\cite{ANTARES_PointSources2010}, as well as a recently developed estimator exploiting the correlation between the energy deposit, $\mathrm{d}E/\mathrm{d}X$, and the primary energy~\cite{ICRC2013_dEdX, ANTARES_NuSpectrum}. Both provide very similar results. As shown in Figure~\ref{fig:Performance_Energy}, the $n_\mathrm{Hit}$ energy estimator shows a slightly better performance for weak sources and is therefore retained for the final analysis.

Pseudo-experiments as described below are used to determine the optimal size of the angular steps $\Delta \Omega$. Increasing the number of angular steps enhances the angular resolution of the method but degrades the sensitivity due to the increasing number of trials ({\it look-elsewhere-effect}~\cite{LEE}). Taking into account the median angular resolution of $0.5^\circ$~\cite{ANTARES_PointSources2010}, an optimum has been found for angular steps of about $0.1^\circ$.

\begin{figure*}[!t]
\centerline {
   \includegraphics[width=0.75\textwidth]{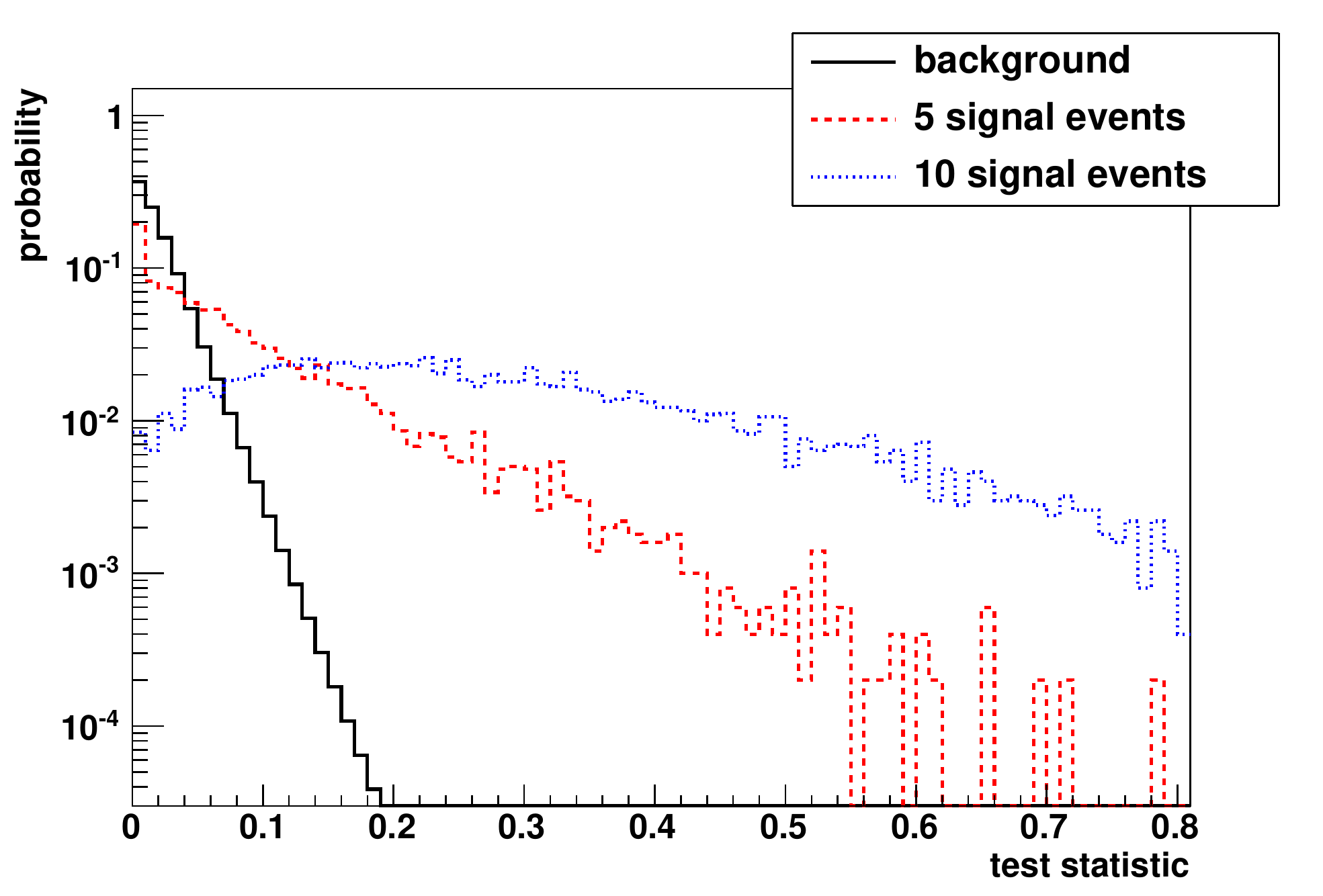}
}
\caption{Normalized distribution of the maximum value of the test statistic derived from pseudo-experiments with scrambled data (black solid line). The red dashed (blue dotted) line corresponds to pseudo-datasets including a point-like source of 5 (10) events.}
   \label{fig:PE}
 \end{figure*}
\noindent%

\subsection{Reference autocorrelation distribution and comparison with data}\label{sec:reference}
To detect structures in the sky distribution of the selected events, a reference autocorrelation distribution to compare with is needed. This reference is determined by scrambling the data themselves, a method which allows the reduction of systematic uncertainties potentially introduced by the use of Monte Carlo simulations. The scrambling is performed keeping the pairs of local coordinates (zenith, azimuth) in order to avoid losing information about possible correlations between them. The detection time is drawn randomly from another event within the same detector configuration to keep track of the changing layout of the detector due to its construction and maintenance. This method is applied to all selected events and a randomised sky map naturally reproducing the coverage of the unscrambled ANTARES data is constructed. 

This randomised sky is then analysed in exactly the same way as the data to derive the autocorrelation function. The randomisation process is performed about $10^6$ times and the derived autocorrelation distributions are averaged in order to reduce statistical fluctuations. 

Structures in the sky distribution of the data will show up as differences between the autocorrelation distribution of the data and the reference distribution. The comparison is performed by using the formalism introduced by Li and Ma~\cite{LiMa}. This formalism provides a raw test statistic, $t$, as a function of the cumulative angular scale. As the comparison is performed bin-by-bin and as the scan is made over different angular scales, this result has to be corrected for the corresponding trial factor. To limit the number of trials the scan is performed only up to $25^\circ$, a scale which includes most known extended sources and emission regions.

Finally, the method proposed by Finley and Westerhoff~\cite{FinleyWesterhoff2004} is applied by performing about $10^6$ pseudo experiments in which the autocorrelation distributions of randomised sky maps are compared with the reference distribution. For each simulated map the maximum value of the test statistic is calculated. The obtained distribution is shown as black solid line in Figure~\ref{fig:PE}. To reach high significances the tail of the obtained distribution is fitted and extrapolated with an exponential function. The final p-value of the analysis is then calculated as the probability to obtain the same or a higher value of $t$ from these background-only pseudo-experiments.\\
 
\subsection{Performance and sensitivity}\label{sec:pe}

\begin{figure*}[!t]
  \centerline{
  \includegraphics[width=0.49\textwidth]{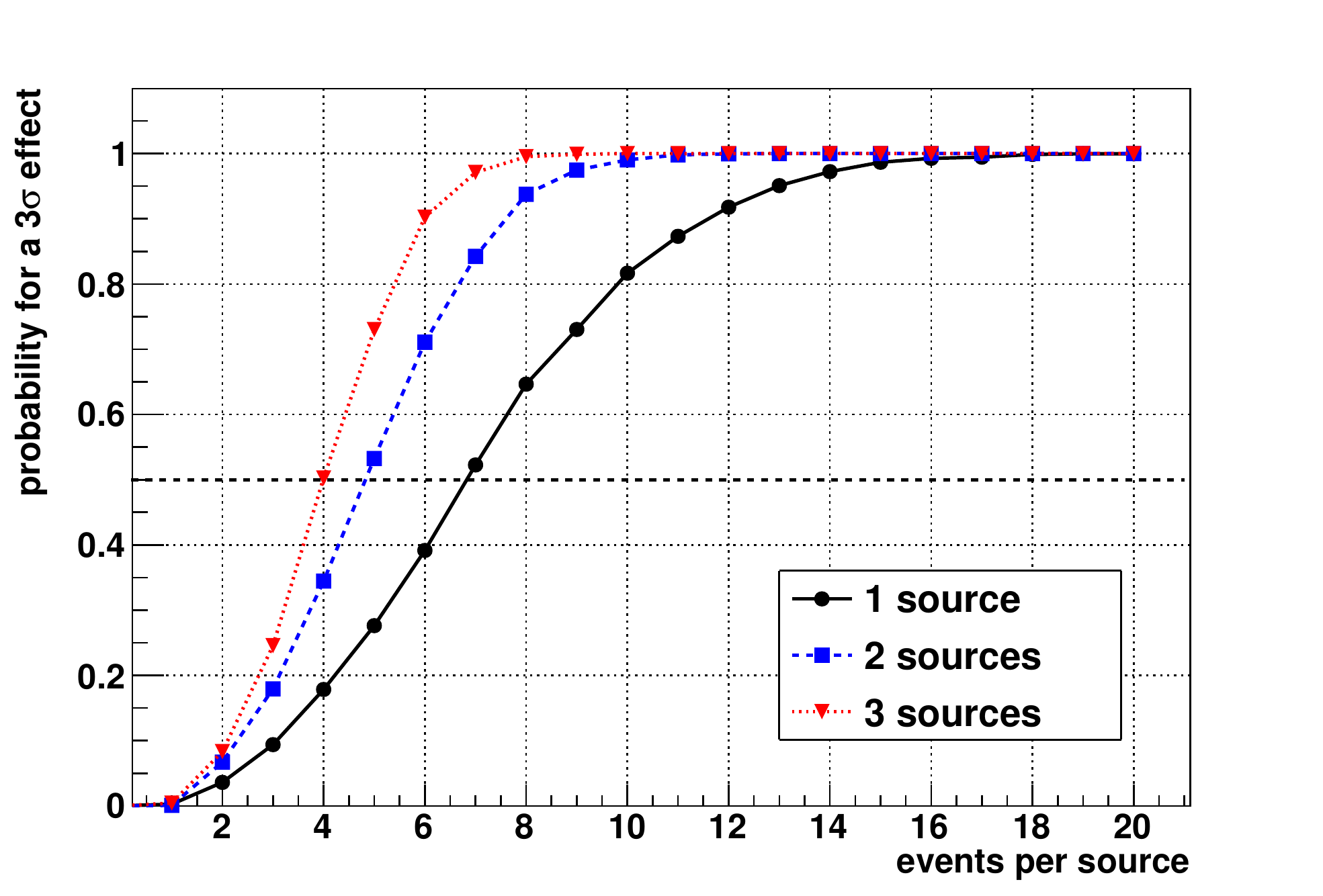}
  \hfill
  \includegraphics[width=0.49\textwidth]{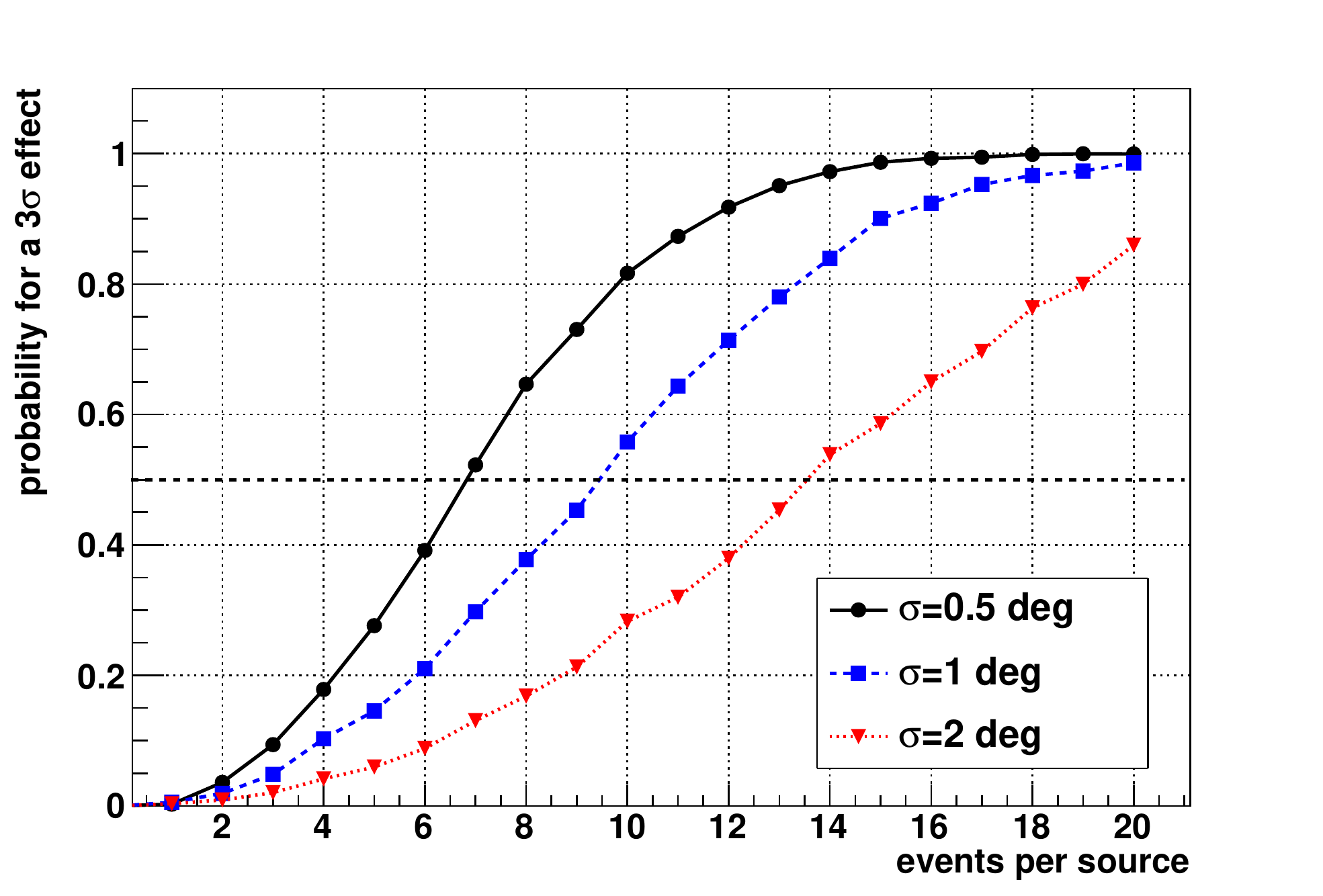}
   }	
  \caption{Probability for a $3~\sigma$ effect using the autocorrelation method exploiting the $n_\mathrm{Hit}$ energy estimator. Left plot: Dependence on the number of injected point-like sources in the visible sky. Right plot: Dependence on the extension of a single source modelled by a 2-dimensional Gaussian.}
  \label{fig:nHitsPerformance}
 \end{figure*}
 
The performance of the algorithm is determined using mock datasets built by scrambling the selected data events as described above. While keeping the total number of events constant, predefined source structures with various sizes and source luminosities are added. The angular resolution of the detector is taken into account by convolving the intrinsic source size with a two dimensional Gaussian with a width of $\sigma=0.5^\circ$. The energy estimater for the injected signal events is drawn randomly from distributions weighted to follow an $E^{-2}$ energy spectrum (see red dotted line in Figure~\ref{fig:nHit}). These mock datasets are then analysed in exactly the same way as described in Section~\ref{sec:reference}. 

Compared to a dedicated likelihood-based search for a point-like excess in the same dataset~\cite{ANTARES_PointSources2010}, the sensitivity of the autocorrelation analysis is slightly worse for a single source. The present method indeed requires about 7 signal events to obtain a $3~\sigma$ detection with a $50~\%$ probability, compared to about 6 events required in the likelihood search. On the other hand, it outperforms the algorithm optimised for the localisation of point-like sources as soon as several weak sources are present, which underlines the complementarity of the two methods. Another advantage is the sensitivity of the autocorrelation method to extended source regions. The performance of the algorithm for both cases is illustrated in Figure~\ref{fig:nHitsPerformance}.\\

\subsection{Autocorrelation results and discussion}
\begin{figure*}[!t]
\centerline {
     \includegraphics[width=0.85\textwidth]{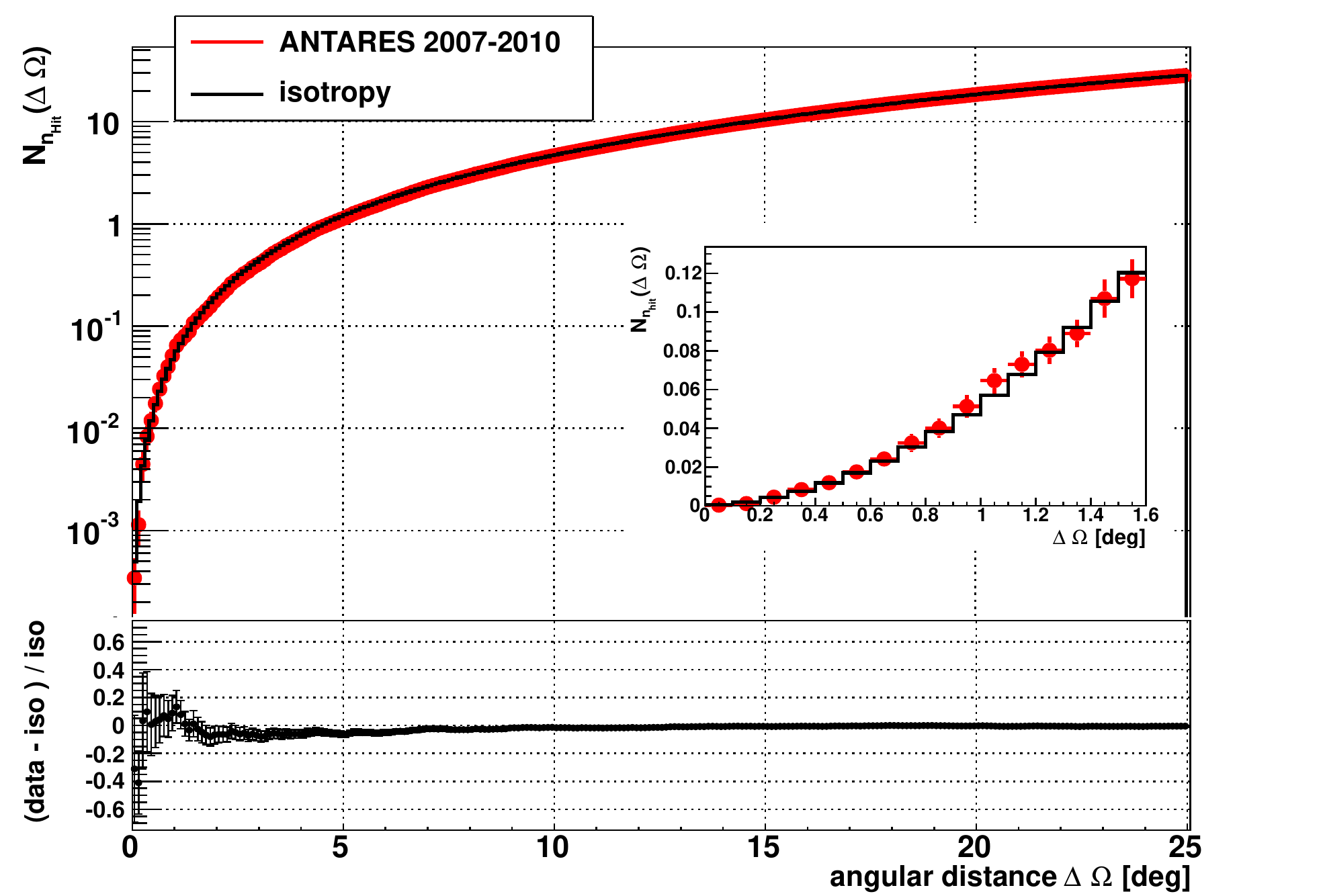}
}
\caption{Cumulative autocorrelation function of data taken with the ANTARES neutrino telescope in 2007-2010. The red markers denote the ANTARES data and the black histogram represents the reference distribution expected for an isotropic dataset. The inset shows an enlarged view for small angular distances and the lower panel depicts the relative difference between data and reference distribution.}
   \label{fig:Sky_AutoCorrelation}
 \end{figure*}
\noindent%

Following Eq.~\ref{equ:autocor}, the improved cumulative autocorrelation analysis using the $n_\mathrm{Hit}$ energy estimator is applied to the 3058 selected neutrino candidate events recorded by the ANTARES neutrino telescope between 2007 and 2010. The obtained distribution is shown as the red markers in Figure~\ref{fig:Sky_AutoCorrelation} and compared with the reference corresponding to the expectation from an isotropic distribution of the arrival directions (black histogram). The maximum deviation between the data and the reference distribution is found for an angular scale $\leq1.1^\circ$. Correcting for the scanning trial factor this corresponds to a p-value of $9.6~\%$ and is therefore not significant. In addition, it is known that the dataset analysed here contains a slight excess of events around (R.A., Dec) = ($-46.5^\circ$, $-65.0^\circ$), where a cluster of 5 events within one degree has been found~\cite{ANTARES_PointSources2010}. This cluster resulted in a $2.2~\sigma$ effect and the region has been studied in detail in Ref.~\cite{ICRC2013_ANTARESExcess}. Replacing these events by randomised events increases the post-trial p-value of the autocorrelation analysis to $35~\%$. Therefore the analysed ANTARES dataset does not contain significant clusters in addition to the small point-like excess that had already been observed in the dedicated search.
 
\section{Two-point cross-correlation with external catalogues}\label{sec:2pt}
One way to improve the sensitivity of searches for sources of high-energy astrophysical neutrinos is to rely on the connection with other messengers. Based on phenomenological source scenarios, observations in certain wavelengths and catalogues of interesting astrophysical objects can provide valuable additional information. This approach is followed here through a first search for a global correlation between neutrinos detected by the ANTARES telescope and high energy gamma rays as well as the matter distribution in the local universe represented by the distribution of galaxies. The latter correlation with extragalactic sources, is complemented by a correlation with a catalogue of massive black holes. A dedicated correlation analysis between ultra-high energy cosmic rays detected by the Pierre Auger Observatory and neutrino candidates recorded by ANTARES has been published in Ref.~\cite{AugerCorrelation}.

For this purpose, the autocorrelation function described in Eq.~\ref{equ:autocor} is extended to measure the two-point cross-correlation between the $N$ neutrino candidates and an external dataset of $n$ astrophysical objects: 
\begin{equation}
\mathcal{N}_\mathrm{p} (\Delta \Omega) = \sum\limits_{i=1}^{N} \sum\limits_{j=1}^{n} w_{i} \cdot \hat{w}_{j} \cdot \left[ 1 - H(\Delta \Omega_{ij} - \Delta \Omega) \right], \label{equ:2ptcorr}
\end{equation}
Here, $w_{i}$ denotes the weights derived for each neutrino candidate event as described in Section~\ref{sec:method}. The weights related to the external dataset, $\hat{w}_{j}$, are calculated in a similar way, i.e. by integrating the normalized distribution $f(\hat{x})$  of the discriminant parameter $\hat{x}$: $\hat{w}_{j}= \int_0^{\hat{x}_j} f(\hat{x})\; \mathrm{d}\hat{x} $. The methods for the calculation of the reference distribution expected from an isotropic neutrino dataset, the comparison with the data and the correction for trial factors using pseudo experiments is performed in the same way as described in Section~\ref{sec:autocorr} for the autocorrelation analysis.

\begin{figure*}[!t]
  \centerline{
\includegraphics[width=0.49\textwidth]{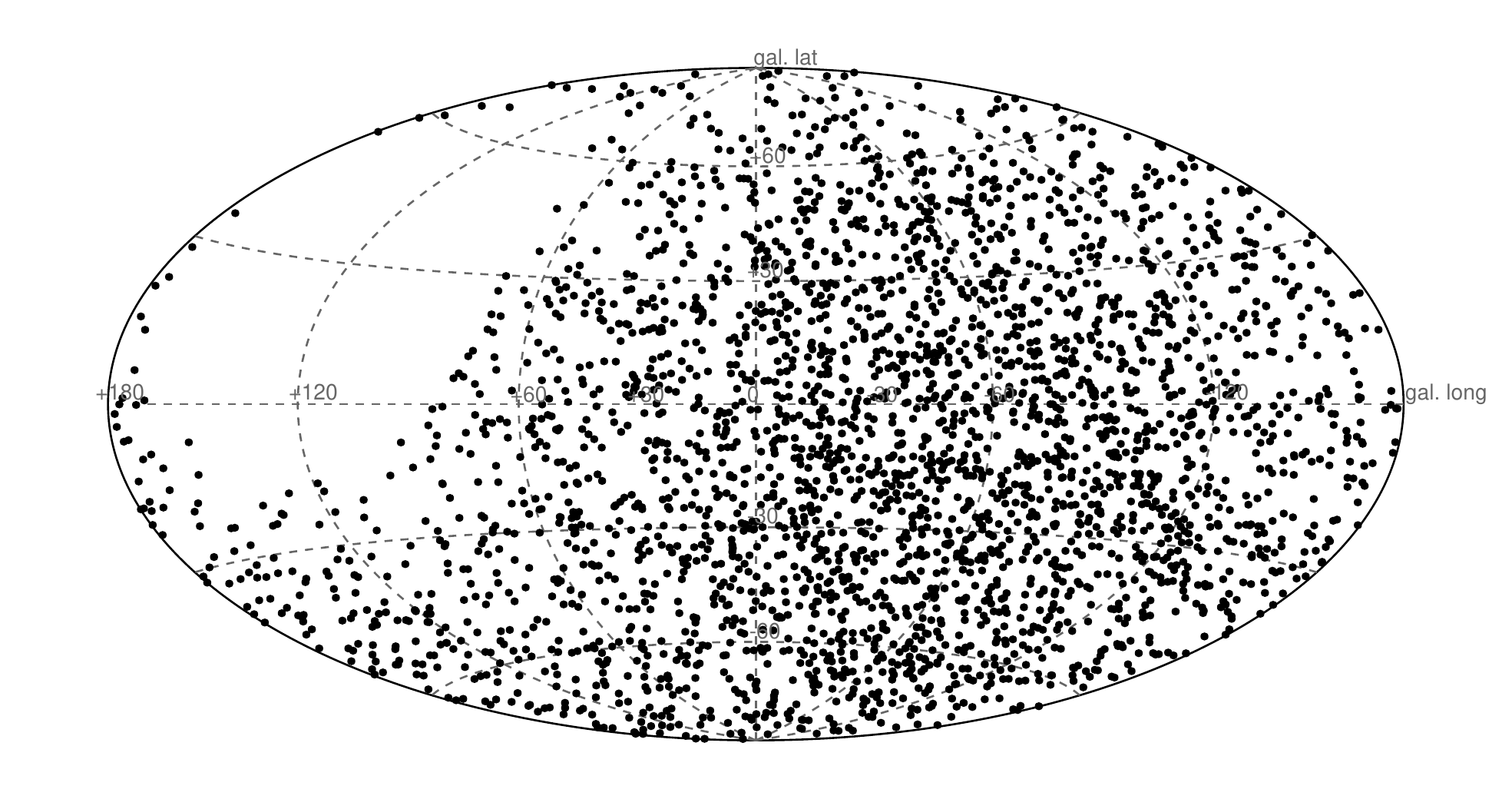}
\hfill
  \includegraphics[width=0.49\textwidth]{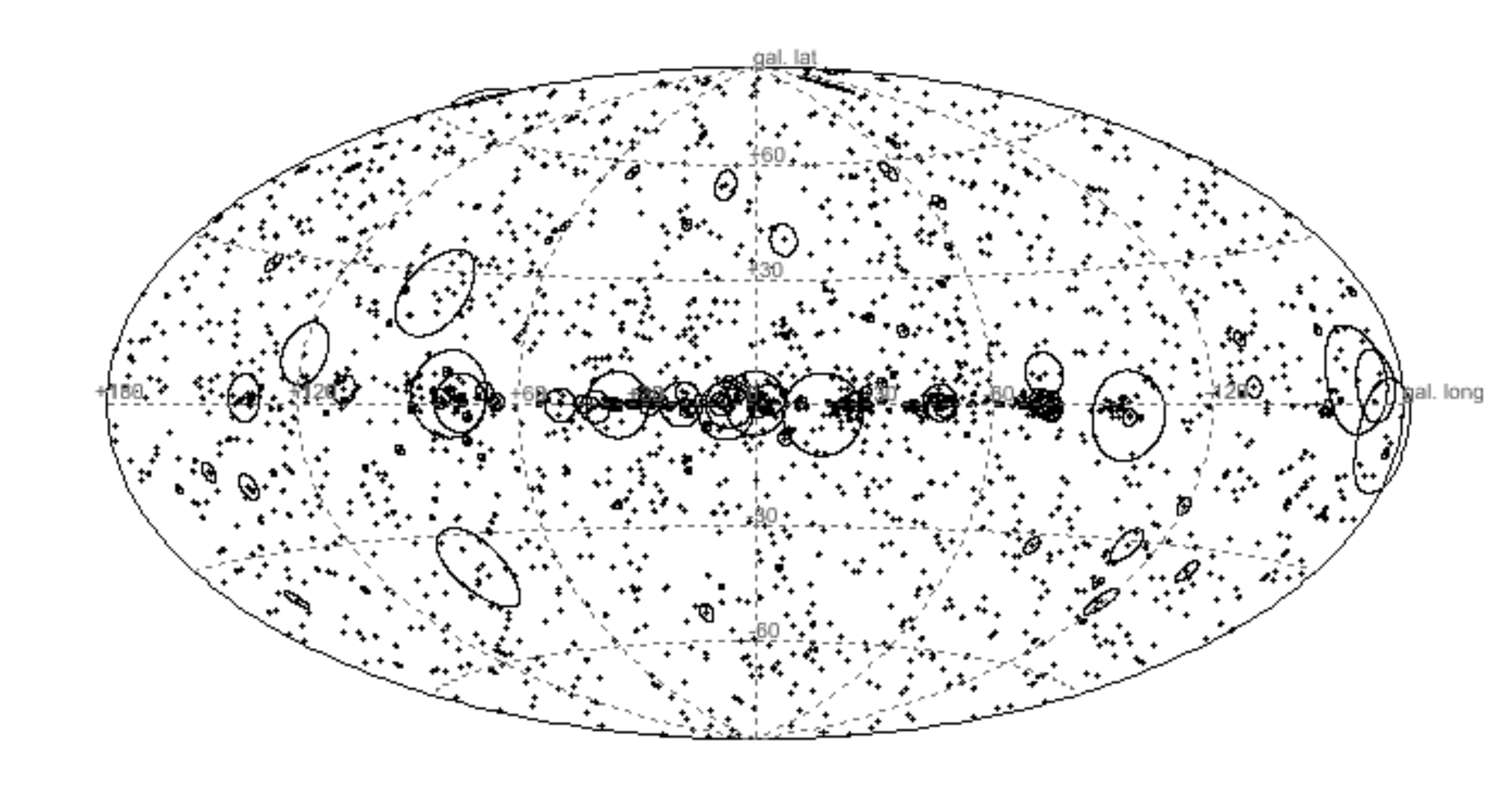}
  }\vspace{-55mm}
  \centerline{
\includegraphics[width=0.49\textwidth]{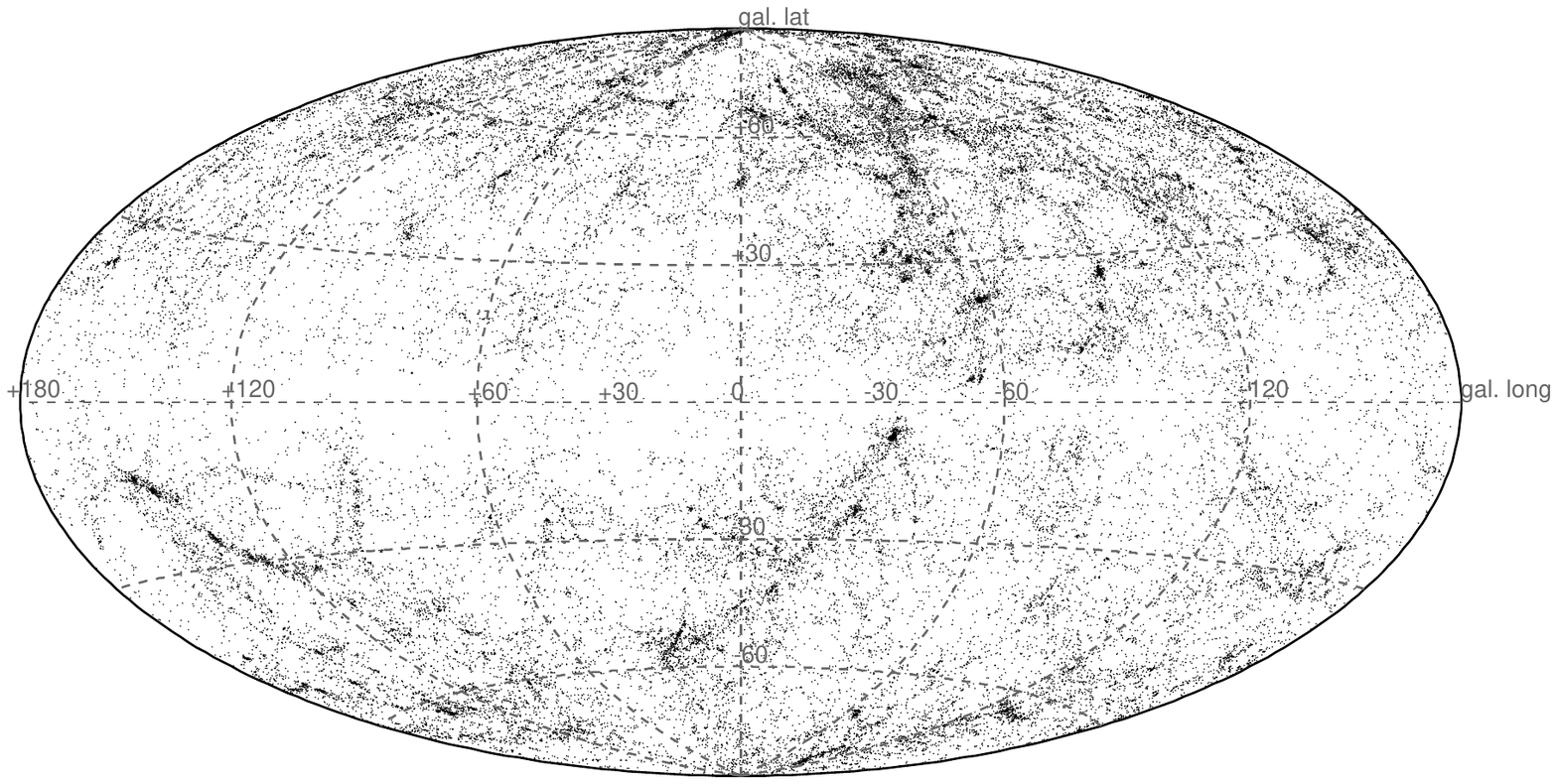}
  \hfill
  \includegraphics[width=0.49\textwidth]{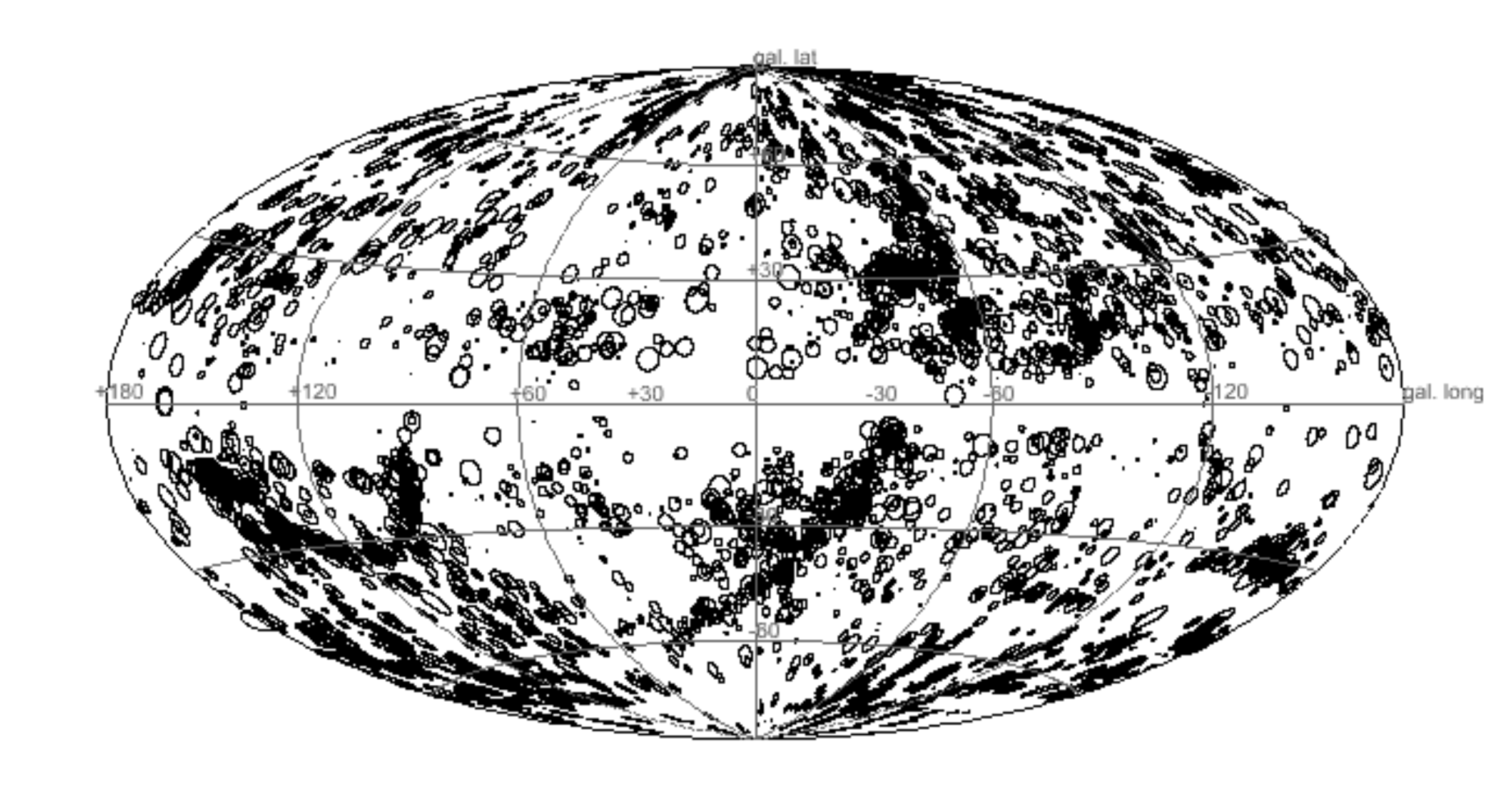}
      }	
  \caption{Upper left plot: Skymap in galactic coordinates of the 3058 selected neutrino candidates used for this analysis~\cite{ANTARES_PointSources2010}. Upper right plot: High-energy gamma ray sources given in the 2FGL catalogue from Fermi-LAT~\cite{2FGL}. The size of the circles indicates the gamma ray flux in the $1-100~\mathrm{GeV}$ energy range. Lower left plot: Galaxies within $100~\mathrm{Mpc}$ as given in the GWGC catalogue~\cite{GWGC}. Lower right plot: Massive black holes as given in~\cite{BlackHoleCatalog}. The size of the circles indicates the mass of the objects.}
  \label{fig:SkyMaps}
 \end{figure*}
 \noindent%
 
\subsection{High-energy gamma rays}
Data from two years of observation of high-energy gamma rays with the Fermi-LAT satellite is used to compile the 2FGL point source catalogue~\cite{2FGL}. It is shown in the upper right plot of Figure~\ref{fig:SkyMaps}. The full catalogue, containing 1873 gamma ray sources, is used for a two-point correlation analysis with the selected ANTARES neutrino candidates. It should be noted that a small subset of these sources are also included in the candidate list used in a dedicated search for point-like sources~\cite{ANTARES_PointSources2010}. Each 2FGL source is weighted with its gamma ray flux $1-100~\mathrm{GeV}$ as given in the Fermi catalogue and the ANTARES events are weighted based on the $n_\mathrm{Hit}$ energy estimator. The minimum post-trial p-value of $68~\%$ is found for angular scales smaller than $0.6^\circ$.\\

\subsection{The local universe}
The locations of the yet unknown cosmic ray accelerators are likely correlated with the matter distribution in the local universe. To exploit this connection, the 'Gravitational Wave Galaxy Catalogue' (GWGC) which provides a rather complete set of galaxies within a distance of $D<100~\mathrm{Mpc}$, is used as description of the local extra-galactic matter distribution~\cite{GWGC}. Their distribution is shown in the lower left plot Figure~\ref{fig:SkyMaps}. Assuming the simplest case of equal neutrino luminosity from all given 53295 galaxies, a $D^{-2}$ weighting for the galaxies and the $n_\mathrm{Hit}$ neutrino weights for the neutrino candidates are used. The two-point correlation analysis finds the most significant clustering at scales smaller than $0.3^\circ$ with a post-trial p-value of $96~\%$.\\

\subsection{Massive black holes}
A refinement of this largely unbiased analysis is the introduction of selection criteria that favour cosmic ray accelerator candidates among the neighbouring galaxies. For example, the sub-class of galaxies housing massive black holes at their centers has been discussed as efficient accelerators of cosmic rays up to ultra-high energies (for a summary of proposed acceleration sites, see e.g. Ref.~\cite{Stanev-UHECR}). Here this scenario is exploited by searching for correlations between the neutrino candidates detected by the ANTARES telescope and massive black holes given in Ref.~\cite{BlackHoleCatalog} (see lower right plot of Figure~\ref{fig:SkyMaps}). The weighting the 5894 objects in the catalogue according to their mass reflects the energetics of the black hole systems and thus their acceleration power. Again, the neutrino events are weighted using the $n_\mathrm{Hit}$ estimator. The minimum post-trial p-value of $56~\%$ is found for angles smaller than $8.6^\circ$.

\section{Summary}
In the search for the sources of high-energy cosmic rays, the detection of astrophysical neutrino sources may play a crucial role. Here a search for intrinsic clustering of data recorded with the ANTARES neutrino telescope is presented. This analysis uses an improved two-point correlation technique exploiting an estimate of the energy of the neutrino candidates. The arrival directions of the selected neutrino candidates neither show evidence for clustering of events on top of the isotropic distribution expected for the background of atmospheric neutrinos, nor correlate with catalogues of gamma rays, nearby galaxies or massive black holes.


\acknowledgments
The authors acknowledge the financial support of the funding agencies: Centre National de la Recherche Scientifique (CNRS), Commissariat \`a l'\'energie atomique et aux \'energies alternatives (CEA), Commission Europ\'eenne (FEDER fund and Marie Curie Program), R\'egion Alsace (contrat CPER), R\'egion Provence-Alpes-C\^ote d'Azur, D\'e\-par\-tement du Var and Ville de La Seyne-sur-Mer, France; Bundesministerium f\"ur Bildung und Forschung (BMBF), Germany; Istituto Nazionale di Fisica Nucleare (INFN), Italy; Stichting voor Fundamenteel Onderzoek der Materie (FOM), Nederlandse organisatie voor Wetenschappelijk Onderzoek (NWO), the Netherlands; Council of the President of the Russian Federation for young scientists and leading scientific schools supporting grants, Russia;
National Authority for Scientific Research (ANCS), Romania; Ministerio de Ciencia e Innovaci\'on (MICINN), Prometeo of Generalitat Valenciana and MultiDark, Spain; Agence de l'Oriental and CNRST, Morocco. We also acknowledge the technical support of Ifremer, AIM and Foselev Marine for the sea operation and the CC-IN2P3 for the computing facilities.


\end{document}